\documentclass[12pt,preprint]{aastex}

\newcommand{\chandra}{\emph{Chandra} }
\newcommand{\Chandra}{\emph{Chandra} }
\date{\today}

\begin{document}

\title{Chandra high resolution X-ray spectroscopy of AM~Her}

\author{V. Girish, V.R. Rana \and K.P. Singh}
\affil{Department of Astronomy \& Astrophysics, Tata Institute of Fundamental
Research, Mumbai - 400005, INDIA}
\email{giri@tifr.res.in,vrana@tifr.res.in,singh@tifr.res.in}

\begin{abstract}

We present the results of high resolution spectroscopy of the prototype
polar AM~Herculis observed with  \Chandra \emph{High Energy Transmission
Grating}.  The X-ray spectrum contains hydrogen-like and helium-like lines
of Fe, S, Si, Mg, Ne and O with several Fe L-shell emission lines. The
forbidden lines in the spectrum are generally weak whereas the
hydrogen-like lines are stronger suggesting that emission from a
multi-temperature, collisionally ionized plasma dominates.  The helium-like
line flux ratios yield a plasma temperature of 2~MK and a plasma density
1--9 $\times 10^{12}\ cm^{-3}$, whereas the line flux ratio of \ion{Fe}{26}
to \ion{Fe}{25} gives an ionization temperature of 12.4$^{+1.1}_{-1.4}$
keV.  We present the differential emission measure distribution of AM~Her
whose shape is consistent with the volume emission measure obtained by
multi-temperature APEC model. The multi-temperature plasma model fit to the
average X-ray spectrum indicates the mass of the white dwarf to be $\sim
1.15$~M$_\sun$.

From phase resolved spectroscopy, we find the line centers of \ion{Mg}{12},
\ion{S}{16}, resonance line of \ion{Fe}{25}, and \ion{Fe}{26} emission
modulated by a few hundred to 1000 km~s$^{-1}$ from the theoretically
expected values indicating bulk motion of ionized matter in the accretion
column of AM~Her.  The observed velocities of \ion{Fe}{26} ions are close
to the expected shock velocity for a 0.6 M$_\sun$ white dwarf.  The
observed velocity modulation is consistent with that expected from a single
pole accreting binary system.

\end{abstract}
	\keywords{novae, cataclysmic variables -- stars:individual-(AM Her) --
	X-rays: binaries}

\section{ Introduction }
\label{sec-intro}

Polars are a class of magnetic cataclysmic variables that are synchronously
locked, close interacting binary systems with a highly magnetized white
dwarf primary. The high magnetic field that can range from 10~MG to 230~MG
prevents  the formation of an accretion disk and channels the accreting
matter along the field lines to the surface of the white dwarf. The slowing
down of supersonic accreting matter forms a shock near the surface of the
white dwarf.  The shock heats up the in-falling matter to very high
temperatures ($\sim 10^8$ K) and the hot plasma in the post shock region
cools off by emitting hard X-rays via thermal   bremsstrahlung process. A
part of these hard X-rays heat up the surface of the white dwarf which in
turn emits soft X-rays.

AM~Her, is the prototype of polars at a distance of 91~pc
\Citep{gansicke95}.  According to \Citet{young81}  AM~Her has an orbital
period of 3.1~hours and consists of a white dwarf of mass $0.39\ M_\sun$
and radius $0.26\ R_\sun$, and an M dwarf secondary with a mass of $0.26
M_\sun$.  However, various other estimates for the mass of the white dwarf
reported in the literature quote values in the range  of 0.5 M$_\sun$ to
1.22 M$_\sun$ \Citep{gansicke98,wu95,mukai87,mouchet93,cropper98}.

High resolution X-ray spectroscopy with \Chandra's High Energy Transmission
Grating (HETG) allows us to resolve many spectral lines which in turn allow
us to study the X-ray emission processes, and temperature, density and
ionization properties of the X-ray plasma. The high resolving power of
\Chandra can also help us to determine the Doppler shifts of emission
lines, which can be used to study the kinematics of X-ray emitting region
in Cataclysmic Variables.

Here, we present the results of high resolution X-ray spectroscopy of
AM~Her based on the analysis of archival \Chandra data taken with HETG
\Citep{canizares05}.  The organization of the paper is as follows.  In
\S\ref{sec-obs}, we describe the observations, data analysis and the search
for the presence of Doppler shifts of different emission lines, discussion
in \S\ref{sec-disc},  and conclusions in \S\ref{sec-concl}.

\section{ Observations and Analysis }
\label{sec-obs}

\Chandra observed AM Her on 2003 August~15
(\dataset[ADS/Sa.CXO#obs/03769]{ObsId 3769}) using HETG for 94~ks. The data
were continuous and covered $\sim$8.5 orbital cycles.  The HETG
observations were taken in combination with the \emph{Advanced CCD Imaging
Spectrometer} (ACIS; \Citealt{garmire03}) in the faint mode.  The HETG
carries two transmission gratings: Medium Energy Grating (MEG) and High
Energy Grating (HEG).  The absolute and relative wavelength accuracy of HEG
are $\pm$0.006 $\AA$ and $\pm$ 0.001 $\AA$ respectively (\Chandra
proposers' observatory guide v.7).  This observation was previously used in
a comparative study of iron K$\alpha$ complex in five of the  magnetic
cataclysmic variables observed by \chandra at that time \citep{hellier04}.

AM~Her generally remains in a high state ($V_m \sim 13$~mag), dropping into
a low state ($V_m \gtrsim 15$~mag) occasionally.  The observations reported
by the  \emph{American Association of Variable star observers} (AAVSO;
\Citealt{mattei80}) near the \Chandra observation date show that the star
was in an intermediate state during the HETG observations ($V_m \sim 14$).

\label{sec-avg}

We used the \emph{Chandra Interactive Analysis of Observations} (CIAO) v3.2
software package for data reduction and the extraction of the spectra, and
XSPEC v12.2 \Citep{arnaud96} for spectral analysis.
The lightcurves  extracted from un-dispersed zeroth order HETG image show
eclipses. There is no X-ray flaring activity in the source or in the
background during the times of
observations, and hence, no data were rejected.

\begin{figure*}[htbp]
	\includegraphics[scale=0.8]{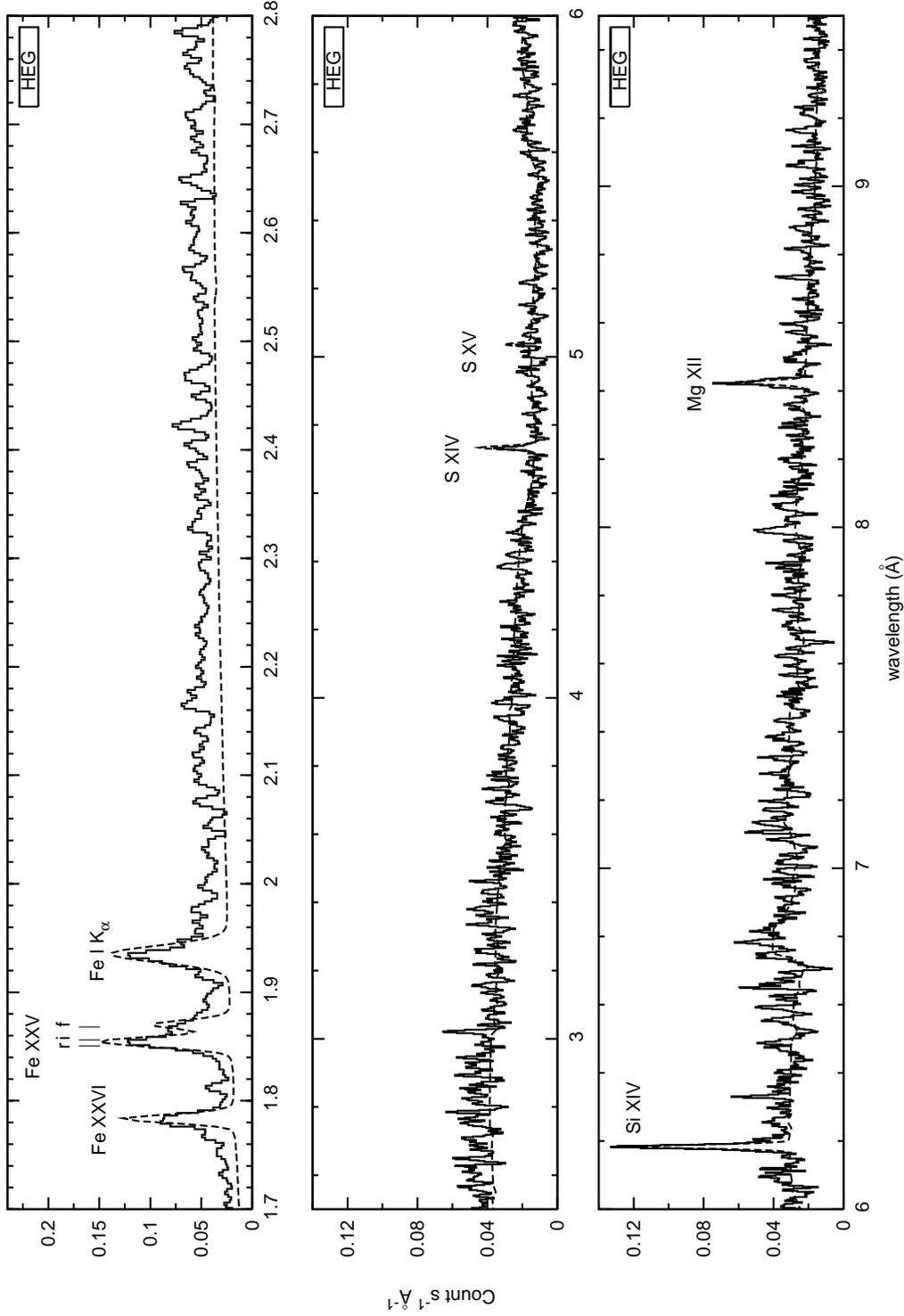}

	\caption{Chandra HETG spectrum of AM Her for the summed +1 and
	-1 orders. Only HEG spectrum is plotted for
	clarity. Data plotted are smoothed by a Gaussian convolution
	of width 0.0025~$\mathrm{\AA}$ HEG.
	The data are shown as histograms and the dashed line
	represents line-by-line fit to the data. Gaussians are used to model
	individual emission lines from various ionized elements and the
	continuum is modelled by an unabsorbed bremsstrahlung model.}

	\label{fig-spectra}
\end{figure*}

\begin{figure*}[htbp]
	\includegraphics[scale=0.8]{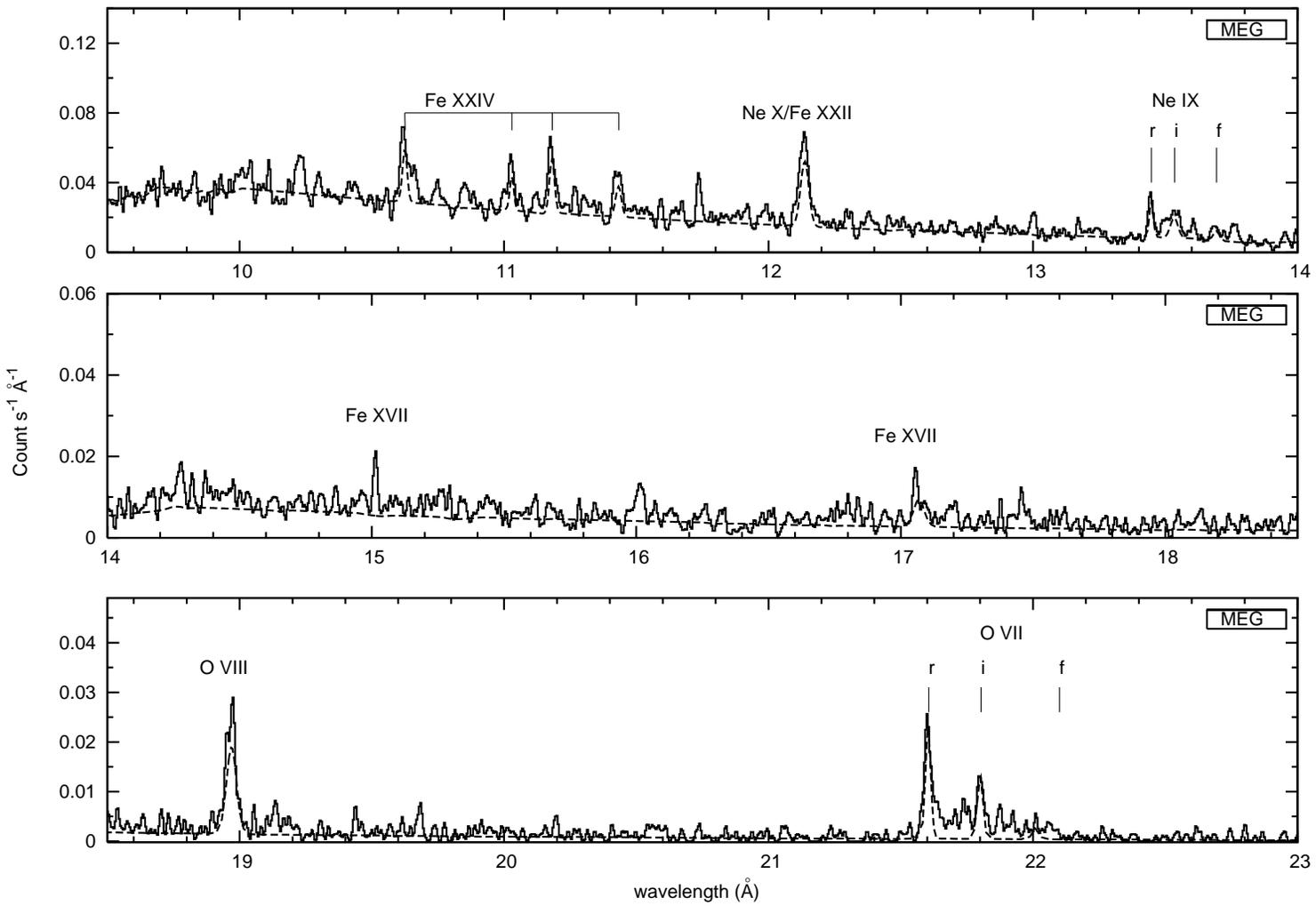}

	\caption{Same as Figure~\ref{fig-spectra}, showing only MEG data. Here
	data are smoothed by a Gaussian convolution of width
	0.005~$\mathrm{\AA}$.}

	\label{fig-spectrab}
\end{figure*}

Using the \emph{tgextract} tool available in CIAO, the spectra for -1 and
+1 orders of HEG and MEG were extracted and summed together using
\emph{add\_grating\_orders} script to improve the statistics. We used the
default binning of 0.0025 $\mathrm{\AA}$ for HEG and 0.005 $\mathrm{\AA}$
for MEG.  Background subtraction is not applied to the data, since the
background counts are less than 1\% of the source counts.  The average
\chandra HEG and MEG spectra of AM~Her are plotted in
Figure~\ref{fig-spectra} and Figure~\ref{fig-spectrab} respectively.  For
clarity, the data were smoothed by a Gaussian of width 0.0025~$\AA$ (HEG)
and 0.005~$\AA$ (MEG) \Citep{david03}.  Several emission lines are seen,
and were identified using \emph{Astrophysical Plasma Emission Database}
(APED; \Citealt{smith01}).  The lines identified are hydrogen-like lines of
\ion{Fe}{26}, \ion{S}{16}, \ion{Si}{14}, \ion{Mg}{12}, \ion{Ne}{10} and
\ion{O}{8}, helium-like triplets of \ion{Fe}{25}, \ion{Ne}{9} and
\ion{O}{7}, fluorescent \ion{Fe}{1} line and several Fe L-shell lines. The
hydrogen-like lines of S, Si, Mg, Ne and O are prominent in the spectrum
while their corresponding helium-like lines are comparatively weaker or
nearly absent altogether. The forbidden lines in hydrogen-like triplets of
different ions are very weak compared to the corresponding resonance and
inter-combination lines. Also, the hydrogen-like and helium-like lines of
Fe are very strong along with a Fe fluorescent line signifying the presence
of both high temperature plasma and photoionization in the spectrum.

\subsection{ Global spectrum analysis }

The AM~Her spectrum is well fitted with four component APEC models, with a
neutral absorber and a partial covering absorber with a covering fraction
of 62\%. The best fit values of temperatures from the multi-temperature
APEC global fitting and their corresponding volume emission measures are
summarized in Table~\ref{tab-apec}.  The multi-temperature APEC models give
a reasonable fit to the spectrum, providing a discrete representation of
the temperature distribution of the plasma. In reality a continuous
distribution of temperature is expected representing  the physical
conditions that exist in the X-ray emitting regions of magnetic cataclysmic
variables (MCVs). Therefore, we have fitted multi-temperature plasma model
to AM~Her spectrum, as described below.

The multi-temperature plasma model uses the calculations of \Citet{aizu73}
to predict the temperature and the density of a magnetically confined hot
plasma in the post-shock region.  Full formulation and assumptions used in
the multi-temperature plasma model is described by \Citet{cropper99}. The
multi-temperature plasma  model was obtained from G. Ramsay via private
communication and added to the XSPEC as a local model. The
multi-temperature plasma model and a partial covering absorption component
along with an absorption due to interstellar matter are used to fit the
average spectrum of AM~Her.  To model the fluorescent Fe line a Gaussian
centered at 6.4~keV is included. The ratio of cyclotron to bremsstrahlung
cooling ($\epsilon_0$) is fixed at 20 corresponding to the observed
magnetic field ($\sim$14~-~22 MG) of AM~Her.  Varying the $\epsilon_0$
value over 10 - 30 does not affect the other spectral parameters
significantly.  The mass accretion rate is fixed to 0.4 g s$^{-1}$
cm$^{-2}$ \Citep{gansicke95} and the radius of the accretion column is
fixed to 10$^{8}$ cm, typical of white dwarves. Elemental abundances are
assumed to be  solar, and the accretion column is divided into	100
vertical grids with a viewing angle of 35$^\circ$. The best fit value for
the mass of the white dwarf is found to be 1.15 $\pm$ 0.05 M$_\sun$ with a
reduced $\chi^2_{min}$ of 0.5 for 2875 dof. The mass obtained by using
multi-temperature plasma model is consistent with the mass determined by
\Citet{cropper98}.  However, in this model authors assumed a
one-temperature flow in post shock region of MCVs in which the electrons
and the ions share the same temperatures.  Recently, \Citet{saxton05}
improved on this and performed two-temperature hydrodynamic calculations
for the post shock flow in accretion column of MCVs.  Two-temperature flow
is expected to occur in MCVs with strongly magnetized ($\geq$20 MG) massive
white dwarf ($\sim$1 M$_\sun$). \Citet{saxton05} have compared the results
of two-temperature calculations with the one-temperature calculations and
found that the white dwarf mass estimated by using two-temperature
hydrodynamic spectral models are lower than those obtained by
one-temperature models. This bias is found to be more severe for massive
white dwarf systems.

\subsection{ Line-by-Line Analysis }

In another method, we used Gaussians to fit the individual emission lines
with the aim of measuring the emission line properties.  This method has
the virtue of not assuming plasma parameters inherent in MEKAL or APEC
models, but to derive the plasma properties from the ratios of the fluxes
estimated by using Gaussians. Both HEG and MEG spectra were fitted
simultaneously. The continuum was estimated by an unabsorbed bremsstrahlung
component of temperature 15.5~keV.  Both the continuum and Gaussians were
folded with the instrumental response. The normalization of the
bremsstrahlung component was frozen after fitting the continuum using line
free regions.  The width of each line was tested for significant non-zero
value by initially varying it for each line.  It was found that the line
widths, in general, were consistent with the resolution of the instrument.
Therefore, the widths were fixed at zero to mimic the unresolved lines,
except for the fluorescent line of Fe where the width was allowed to vary.
The line center energy and the photon flux were allowed to vary for each
line during the final fitting except for few weak lines for which the
line-centers were frozen at the respective theoretical energies.  Cash
statistics was used to estimate the parameter values and confidence ranges.
Though Cash statistics is a better criteria to determine the confidence
range on a best fit parameter when the data bins have few counts, in its
original form, Cash statistics could not provide a goodness of fit similar
to $\chi^2$. XSPEC uses a modified function which provides a
goodness-of-fit criterion similar to $\chi^2$ in the limit of larger
counts. The background counts are not subtracted from source counts while
modelling, as the background counts form only 1\% of the source spectrum
counts. Not subtracting the background counts from the source counts does
not affect our interpretation. The best fit model gives a C-statistic value
of 7747 for 7875 dof.  Table~\ref{tab-avgl} lists the fluxes and the
equivalent widths of emission lines along with their probable
identifications.  The model spectrum is plotted as a dashed line in
Figure~\ref{fig-spectra} and Figure~\ref{fig-spectrab}. The individual
emission lines are also marked in the spectrum.

In the average spectrum, the line-centroids of several lines show a
significant deviation from the predicted wavelength of the lines. Assuming
the observed shifts are real and not an artefact, we interpret the shifts
as the shifts caused due to Doppler effect. A comparison of the observed
line centers with the theoretically predicted line centers thus yields
different Doppler velocities for \ion{Fe}{26}, resonance line of
\ion{Fe}{25}, %fluorescent \ion{Fe}{1} line, \ion{S}{16} and \ion{Mg}{12}.
These Doppler shifts are listed in the second column of
Table~\ref{tab-lines}.

\subsection{ Emission measure analysis of the average spectrum }

The volume emission measure (VEM) is the measure of the ``amount of material''
available in a plasma to produce the observed flux,
which gives us an idea of how the emitting material is distributed
with the emitting temperature.  For an optically thin plasma in collisional
ionization equilibrium,
the relation for VEM can be written as \Citep{griffiths98},
%%%
\begin{equation}
\left[\int n_e\ n_h dV\right] (max) = 4\pi d^2 \frac{f_l}{\bar{G_l}}
\label{eqn-em}
\end{equation}
%%%
where, $f_l$ is observed flux in a line feature, $G_l$ is emissivity ($photons \
cm^{-3} s^{-1}$), $d$ is distance to the source, $n_{e}$ and $n_{h}$ are
the electron and hydrogen densities in cm$^{-3}$, respectively.

\begin{figure*}[htp]
	\includegraphics[angle=-90,scale=0.6]{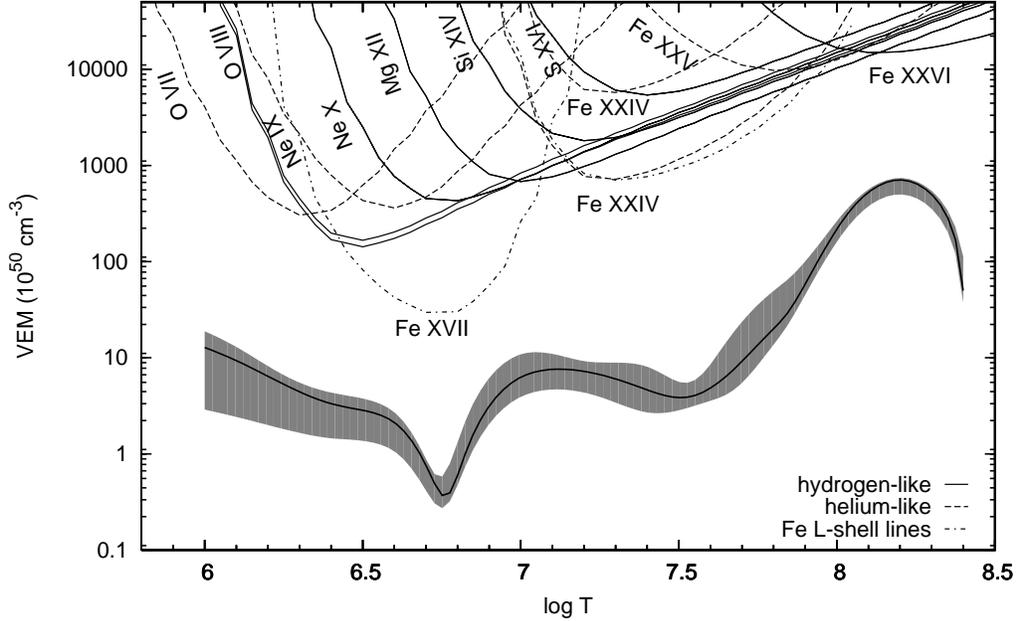}

	\caption{VEM as a function of temperature. The dashed lines, solid
	lines and dash-dot lines corresponds to VEM of helium-like,
	hydrogen-like and Fe-L shell lines respectively.  The thick solid line
	is the DEM derived using MCMC(M) reconstruction technique on a set of
	lines of hydrogen-like, helium-like of different atoms and highly
	ionized Fe line fluxes assuming solar photospheric abundances.  The
	shaded areas  are the 95\% confidence limits corresponding to each
	temperature bin.}

	\label{fig-em}
\end{figure*}

Using Equation~\ref{eqn-em}, we can calculate the maximum emission measure
as a function of temperature for each emission line. The combination of
emission measure of individual lines can be used to constrain the emission
distribution of the source.  The VEM estimated using Equation~\ref{eqn-em}
for different lines are plotted in Figure~\ref{fig-em}. The emissivities of
the lines are taken from APED \Citep{smith01}. The curves are the loci
corresponding to the VEM required to produce the observed flux from an
isothermal plasma as a function of plasma temperature.  At a given
temperature, a point on one of these curves represents a maximum EM that
the plasma can have at that temperature.  The solid lines  in the figure
represent hydrogen-like lines of O, Ne, Mg, Si, S and Fe, dashed lines
correspond to resonance lines of O VII, Ne IX and Fe XXV, whereas the
dash-dot-dash lines represent the VEM of Fe L-shell lines.

The VEM defined in Eqn.\ref{eqn-em} in the logarithmic differential form is
called as Differential Emission Measure (DEM).  The DEM gives a correlation
between the amount of emitting power and the amount of emitting material in
the plasma as a function of temperature.  The DEM was estimated by
performing a Markov Chain Monte-Carlo analysis using a Metropolis algorithm
(MCMC[M]) \Citep{kashyap98} implemented in PINTofALE \Citep{kashyap00} on
the set of sixteen brightest lines in the AM~Her spectrum.  The MCMC(M)
method gives an estimate of emission measure distribution over a
pre-selected temperature region with the DEM defined for each bin. Here, we
used a temperature grid ranging from $\log T = 6.0$ to $\log T = 8.8$ with
$\Delta \log T = 0.2$.  The lower and the upper limit of the temperature
region are choosen to represent the hydrogen and helium-like lines of O
and, hydrogen and helium-like lines of Fe respectively, present in the
spectrum. The intermediate temperatures are constrained by the L-shell
lines of Fe, and helium-like lines of Ne, Mg, S and Si. The reconstructed
DEM is plotted in Figure~\ref{fig-em} with 95\% confidence limits shown as
shaded region.

\subsection{ Phase dependent variation of spectral lines }
\label{sec-phase}

To look for modulations in the continuum and emission lines with the binary
period, the data were folded using the ephemeris for the magnetic phase and
a period of $0.128927$~days as given by \Citet{tapia77}, and
\Citet{young81}.  Two methods were adopted to obtain the phase resolved
spectra. In the first method, the spectra were extracted for five
non-overlapping regions of 0.2 phase interval.  Except for the interval
centered around the orbital phase minima, other intervals had enough number
of counts to estimate the line energy and the photon flux in many of the
bright emission lines.  From these spectra it was observed that while some
lines showed shifts in the line energy with phase, others did not.  The
second method is based on extracting the spectra into continuous
overlapping regions \Citep[See][]{hoggenwerf04}.  In this method, the
spectra were accumulated for twenty phase intervals of width $\Delta\Phi\
=\ 0.25$, with phase intervals starting at $\Phi_{0}$ = 0.0, 0.05, 0.1
$\ldots$, where, $\Phi$ is the orbital phase. Spectra obtained from both
the methods were analysed for their continuum and emission line properties.
The continuum was modelled using a bremsstrahlung after masking the regions
of emission lines identified using APED.

\label{sec-phi_avg}

\begin{figure*}
	\includegraphics[angle=-90,scale=0.5]{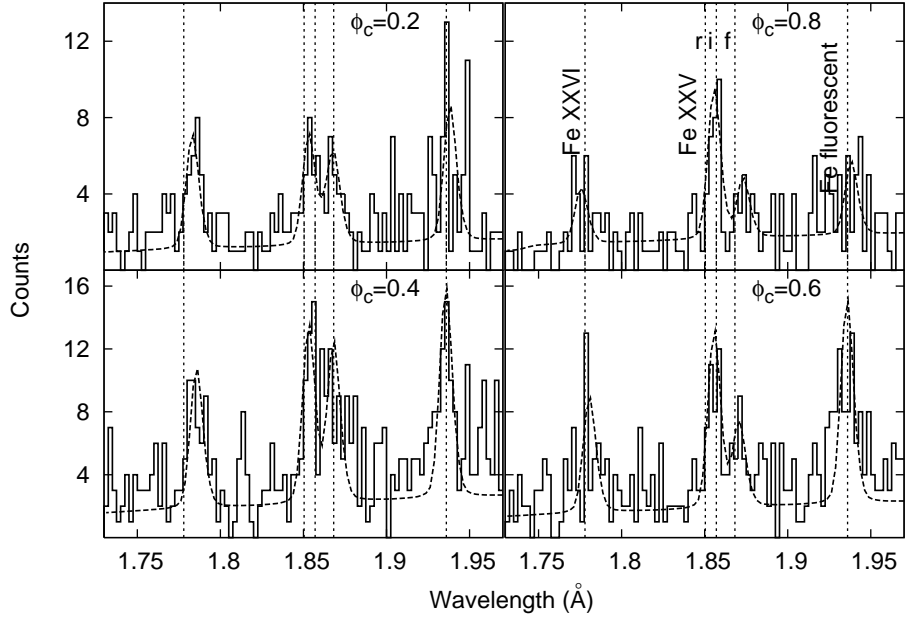}
	\caption{Gaussian fits (dashed lines) to the observed (histogram)
	\ion{Fe}{26}, \ion{Fe}{25} (r,i \& f) and fluorescent Fe lines.  The
	orbital phase is shown in the inset. The width of each phase bin,
	$\Delta \Phi$ = 0.2. }
	\label{fig-felines}
\end{figure*}

\begin{figure}
	\includegraphics[scale=0.8]{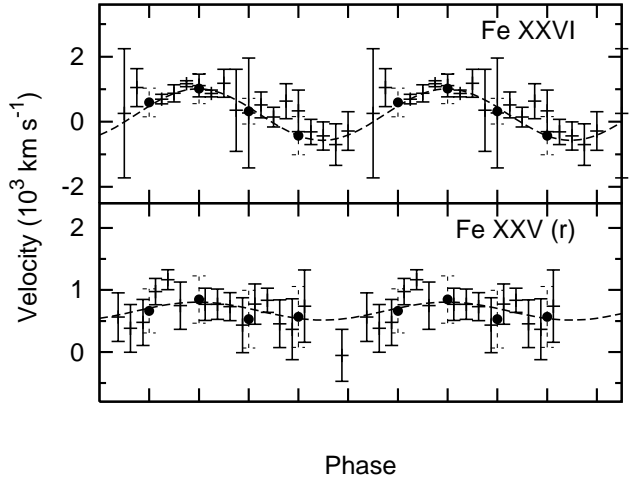}
	\caption{Plot of the velocity shifts as a function of orbital phase for
	the emission lines \ion{Fe}{26} and \ion{Fe}{25} as described in the text.  The filled
	circles correspond to the velocity shifts observed in the
	non-overlapping phase intervals with $\Delta\Phi$ = 0.2.}
	\label{fig-vel_shift}
\end{figure}

\placefigure{fig-felines}
\placefigure{fig-vel_shift}

Different regions of spectra were examined for studying the line emission
properties as a function of phase.  Figure~\ref{fig-felines} shows spectra
for four non-overlapping phase bins in the wavelength region 1.75 -
2.0~$\mathrm{\AA}$ which covers the \ion{Fe}{26} line, \ion{Fe}{25} ($r,\
i,$ and $f$), and the fluorescent line of \ion{Fe}{1}.  The best fit line
energies were compared with the reference line energies to determine the
shifts of different emission lines observed.
The \ion{Fe}{26} line consists of a doublet, Ly$\alpha_1$ and Ly$\alpha_2$
at 6.973 and 6.952 keV respectively. The two lines can not be resolved with
\Chandra HEG resolution of $\sim 35 eV$ at 7~keV. The branching ratio 2:1
of the Ly$\alpha_1$ and Ly$\alpha_2$ yield a line centroid of 6.966 keV
(\Citealt{pike96}; \Citealt{hellier04}). This value was used as a reference
to  determine any shift in the \ion{Fe}{26} line energy.  The stabilizing
transitions of doubly excited ions lead to dielectronic recombination
satellite (DES) lines redward of \ion{Fe}{26} lines.  The \ion{Fe}{26}
lines are mainly affected by a feature at 6.92 keV  whose intensity is
$\sim8\%$ of the \ion{Fe}{26} lines at $5\times10^7\ $K and falls below 5\%
at $8\times 10^7\ $K \Citep{dubau81}.  Hence the contribution of DES lines
to \ion{Fe}{26} in AM~Her can be neglected due to high plasma temperature
of $1.8\times10^8$~K ($\sim$15.6$^{+5.2}_{-3.1}$ keV; the maximum
temperature from 4T APEC fit from Table~\ref{tab-apec}).  The intensities
of DES lines depend both on temperature and intensity of adjacent
ionization states, specially \ion{Fe}{24} and \ion{Fe}{23}. At temperatures
above $3.5\times10^7$~K, the contribution of DES is negligible
\Citep{oelgoetzty01}, and hence, we did not consider the effects of DES
lines. To include the contribution of unresolved $i$ lines and $f$ line, we
used three Gaussians representing the $r$, $i$ and $f$ lines.  For the
calculation of line shifts in \ion{Fe}{25}, we used the rest frame energy
of $r$ at 6.700 keV, $i$ at 6.675 keV and $f$ at 6.636 keV.

The strength of DES lines approximately scales as $Z^4$ and hence, their
effects have been neglected in the analysis of S, Si and Mg. We used 2.623,
2.005 and 1.473 keV as reference line energies for \ion{S}{16},
\ion{Si}{14} and \ion{Mg}{12} respectively.

From the phase resolved spectra, we found that the line centers of
individual lines for different phase bins deviate from the predicted
wavelengths.  The difference between the observed and expected emission
line center is used to calculate the Doppler shift of the emission line
from different phase bins.

The calculated Doppler shifts of emission lines of \ion{Fe}{26} and
\ion{Fe}{25}~(r) as a function orbital phase are plotted in
Figure~\ref{fig-vel_shift}.  Doppler shifts obtained from non-overlapping
phase bins are shown as filled circles.  The shifts from overlapping phase
bins are shown for a better visualization only.  The figure clearly shows a
non-zero shift in the two lines. The amplitude of Doppler shift of
different lines also shows a variation with orbital phase.  The modulation
in the shift in line center with phase is most clearly seen in \ion{Fe}{26}
line as compared to the \ion{Fe}{25} line. In order to quantify the
observed modulation of the Doppler shifts with phase in the above mentioned
lines a sinusoid+constant model was fitted to the data using $\chi^2$
minimization.  To calculate the un-modulated velocity shift, we fit a
sinusoid with non-zero mean to the observed modulation of all lines using
non-linear least square fitting.  We have also checked for the presence of
any Doppler shifts in fluorescent Fe, \ion{S}{16} and \ion{Mg}{12} lines.
The mean Doppler velocity determined from the average spectrum of different
lines are summarized in the second column of Table~\ref{tab-lines}.  The
third and the fourth columns of Table~\ref{tab-lines} list a constant
velocity shift and the amplitude of the sinusoid, respectively.  The
sinusoid best fit for different lines are superposed over the doppler
shifts as dashed lines in Figure~\ref{fig-vel_shift}.  For \ion{Fe}{26} we
get  a semi-amplitude of 790 $\pm$ 40 km~s$^{-1}$ with a mean of 220
$\pm$26~km~s$^{-1}$. Just to compare, we fit a constant model to the
velocity shifts and list them in column five of Table~\ref{tab-lines}.  A
constant fit to the \ion{Fe}{26} data gives a velocity shift 565$\pm$195
km~s$^{-1}$, but the fit is considerably worse with a $\chi^2_\nu = 3.29$
as compared to a $\chi^2_\nu = 0.02$ for the sinusoid fit.
The shifts in the \ion{Fe}{25} (r), \ion{S}{16} and \ion{Mg}{12} are
however well fit with a constant velocity shift of 770~$\pm$~75~km~s$^{-1}$
($\chi^2_\nu=0.06$), 500$\pm$160 km~s$^{-1}$ ($\chi^2_\nu=0.79$), and
280$\pm$195 km~s$^{-1}$ ($\chi^2_\nu=0.15$) respectively. The $\chi^2_\nu$
values for the constant fits and sinusoid fits are similar. The fluorescent
Fe line shows no detectable shift, with $2\sigma$ upper limit of
400~km~s$^{-1}$.  The \ion{Si}{14} line is too weak to determine the
parameters accurately, therefore, no constraints are put on the shift of
this line.

Although no modulation is apparent in the velocity shifts of the
\ion{Fe}{25} (r), \ion{S}{16}, and \ion{Mg}{12}, the error bars on the
shifts of their lines can not rule out the possibility of modulation being
present at some level. To estimate the level of modulation, we have force
fitted a constant plus a sinusoid component similar to the one used for
\ion{Fe}{26}.
The semi-amplitude of the sinusoid and the constant value for the velocity
shifts of \ion{Fe}{25}, \ion{S}{16} and \ion{Mg}{12} are also listed in
Table~\ref{tab-lines}.  The amplitude of modulation in these lines are of
the order of the error bars.

\section{ Discussion }
\label{sec-disc}

\subsection{ Average X-ray spectrum }

Ratios of the line fluxes of helium-like lines provide  good diagnostics of
density and temperature of the line forming regions
\Citep{gabriel69,porquet00}.  The transition from the excited
$^1\mathrm{P}_1, ^3\mathrm{P}_1\, \mathrm{and} ^3\mathrm{S}_1$ levels to
the ground level $^1\mathrm{P}_1$ forms the three strongest lines of helium
triplets: the resonance line (r), the intercombination line (i) and the
forbidden line (f) respectively.  The analytical relations between the
electron density $R(n_e)$ and temperature $G(T_e)$ are, $R(n_e)  =
{f}/{i}$ and $G(T_e)  =  {(f+i)}/{r}$ \Citep{gabriel69,bluementhal72}.

The He-like triplets of oxygen and neon are used as the diagnostics of low
temperature regions. As we could not resolve the He-like lines of Fe, we do
not use \ion{Fe}{25} as a plasma diagnostic.  We can only put an upper
limit on the fluxes of `f' lines of \ion{Ne}{9} and \ion{O}{7} and hence,
we could only estimate the upper limits of the two ratios.  The helium-like
lines of Si, S and Mg are too weak in the spectrum and hence do not allow
us to put any constraints on their G and R ratios.  Using the values listed
in Table~\ref{tab-avgl}, we get $R < 0.48 $ and $G < 0.76 $ for oxygen
triplet and $R < 0.5$ and $G < 0.8$ for neon triplet, as 3$\sigma$ upper
limits.

Comparing the measured $G(T_e)$ values of \ion{O}{7} and \ion{Ne}{9} with
the theoretical relation between $G(T_e)$ and electronic temperatures
\Citep{porquet00} suggests a temperature greater than $\sim 2MK$. This
temperature agrees with the lowest temperature of 0.13~keV (1.5~MK) of the
four APEC components.  Similarly comparing the measured values of $R(n_e)$
with theoretically predicted values for a collisional dominated hybrid
plasma implies a density greater than $2\times 10^{12}\ cm^{-3}$.  Since
the temperature and density of the O and Ne helium-like line emitting
regions are very close, we assume that the emission of the two triplets
takes place very close to each other.  However, it should be noted that
the presence of UV radiation fields can also mimic high densities as the
transition from $f$ to $i$ can also be triggered by UV photons. Strong UV
radiation is known to be present in AM~Her \Citep{wesemael80,szkody82}.

The measured ratio of H- to He-like line intensities can be used to
constrain the ionization temperature of the emitting plasma. We have used
Fe ions to constrain the maximum ionization temperature in the post-shock
region. A value of I(\ion{Fe}{26})/I(\ion{Fe}{25}) = 1.22
$^{+0.27}_{-0.21}$ suggests an ionization temperature of
12.4$^{+1.1}_{-1.4}$~keV \Citep{mewe85}.  This matches well with the
maximum electron temperature 15.6$^{+5.2}_{-3.1}$~keV of AM~Her obtained
from 4-T APEC global fit.  The agreement with the continuum temperature and
ionization temperature is consistent with the results obtained using
moderate energy resolution \emph{ASCA} data \Citep{ishida97}.

Volume emission measure analysis using individual lines shows that the
plasma has a range of temperature from 1.5--100~MK.  The absence of the
\ion{N}{6} He-like triplet indicates an VEM distribution dominated by high
temperatures.  We obtain the lowest VEM from \ion{Fe}{17} with
2.96$\times10^{50}$~cm$^{-3}$  at $\log T=$ 6.8 ($\sim$6~MK), whereas the
peak VEM of 9.73$\times10^{53}$~cm$^{-3}$ from \ion{Fe}{26} is at $\log T=$
8.2 ($\sim$63~MK).  We also estimated the differential emission measure
using the fluxes of sixteen strong lines. We obtain a well constrained
DEM(T) between $\log T = 6.0$ and $\log T = 8.2$.  The minimum of the DEM
is around $\log T = 6.6-6.8$ and there is an indication of a DEM maximum
around $\log T = 8.2$, but, the lack of sufficient number of lines in this
region makes it ambiguous. Though the trial DEM ranges from $\log T = 6.0$
to $\log T = 8.8$, the lack of spectral information beyond $\log T = 8.2$
makes it un-realistic.  The 4-T APEC model provides only an approximate
discrete representation of the continuous VEM distribution. The temperature
distribution thus obtained is consistent with that determined using
individual lines.  The temperature with the lowest VEM/DEM determined by
the former method agrees with the lowest VEM  of 4-T model.

\subsection{ Phase resolved spectroscopy }

According to the standard models of accretion in polars, supersonic matter
in an accretion column forms a shock near the surface of the white dwarf
heating up the post shock region to very high temperatures ($10^8$ K) thus
emitting hard X-rays. As the post shock matter falls toward the white dwarf
surface, it cools down while getting slowed down.  A phase dependent change
in the spectral shape of the continuum  is consistent with a non-uniform
temperature distribution expected in the accretion column of polars.  The
central energies of emission lines of different elements, particularly
hydrogen-like ions of Mg, S and Fe appear to be shifted from their
theoretically expected values, implying possible bulk motion in the
post-shock region of AM~Her. The velocity shift increases as the atomic
number increases from Mg to Fe.  The observed gradient  in the velocities
of different ionization states of different elements can be interpreted as
a signature of cooling and the slowing down of the matter.

The physical conditions in the accretion column can be described by
velocity (v), temperature (T), density ($n_e$), height and size of the
accretion column (h, r). At the shock region, the values of these
parameters depend only on the mass, radius and the accretion rate of the
white dwarf \Citep{aizu73}.  The velocity shift observed in emission lines
of different ionization states of different elements seen in the \Chandra
HEG spectra of AM~Her gives us a unique tool for the determination of the
structure of the accretion column of AM~Her, using the relationships among
the above listed parameters as given by \Citet{aizu73}.

\begin{figure}
	\includegraphics[scale=0.8]{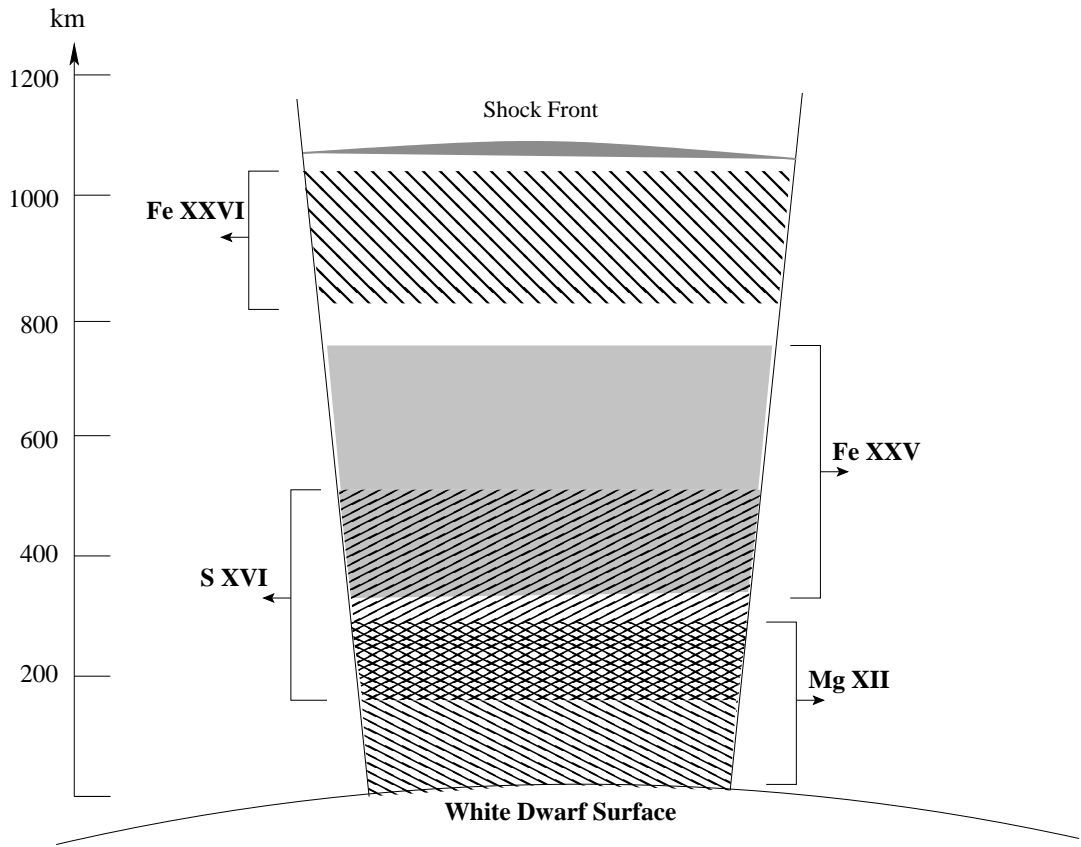}

	\caption{Schematic representation of positions (in km) of  different
	line emitting regions in the accretion column of AM Her. The emitting
	regions are labelled.  The width (in km) of individual emitting region
	corresponds to the errors shown in Table~\ref{tab-lines}.}

	\label{fig-dist}
\end{figure}

The accretion column parameters in the shock region, calculated using
different mass estimates of AM~Her ranging from 0.39--0.91~M$_\sun$
reported in the literature, and an accretion rate of $\dot{m}$ =
0.4~g~cm$^{-2}$~s$^{-1}$ \Citep{gansicke95} are listed in
Table~\ref{tab-shock}. The radius of the white dwarf for each mass value
was calculated using the mass-radius relation for white dwarfs
\Citep{nauenberg72}, and found to be in the range of 0.016--0.009~$R_\sun$.
Using these values we estimated the values of parameters like shock
velocity, $v_{sh}$, shock temperature, kT$_{sh}$ and shock height, $h$, as
listed in Table~\ref{tab-shock} \Citep[See Appendix A,][for the relations
of shock parameters with white dwarf mass, radius and accretion
rate]{terada01}.  We assumed the solar abundances for the elements and mean
molecular weight of the in-falling gas, $\mu$ = 0.615, which corresponds to
a mixture of hydrogen and helium in the ratio of 0.7 and 0.3 respectively.
The calculated $v_{sh}$ for M$_{wd}\ <$~0.6 M$_\sun$ is much less than the
observed velocity shift of the \ion{Fe}{26} line.  Thus, assuming the
observed velocity of \ion{Fe}{26} as equal to $v_{sh}$, we obtain the  mass
of the white dwarf to be $\sim 0.6$~M$_\sun$.  This also puts some
constraints on the geometry of the accretion column of AM~Her by providing
corresponding values of height and radius of the shock region.  For a
0.6~$M_\sun$ white dwarf, we estimated the heights at which different
emission lines are emitted in the accretion column and list them in the
sixth column of Table~\ref{tab-lines}.  Figure~\ref{fig-dist} shows the
schematic of the distribution of different line emitting regions in the
accretion column from the white dwarf surface.  Different shades/patterns
in the figure correspond to different emission lines labelled in the figure
and the width of each region corresponds to 90\% confidence levels of the
height of the emitting region for the particular ion.

A rough estimate about the upper limit on the white dwarf mass can be
obtained using the fact that the heights at which different emission lines
are emitted in the accretion column depend on the mass of white dwarf.
\Citet{wu01} have investigated the ionization structure of the post shock
region of MCVs and predicted line emissivity profiles for \ion{Fe}{26} and
\ion{Fe}{25} lines for white dwarf masses of 1.0~M$_\sun$ and 0.5~M$_\sun$
assuming different magnetic field strengths ($\epsilon_0 = 0, 1, 10, 100$).
They show that highly ionized Fe lines are emitted close to the shock front
for a low mass white dwarf with high magnetic fields, whereas they are
formed well below the shock front for high mass white dwarfs. Though the
emissivity structure depends on the white dwarf mass, the ratio of these
heights will be independent of the mass of the white dwarf. Based on the
observed velocities for \ion{Fe}{26} and \ion{Fe}{25} line emitting
regions, we calculate the ratio of the heights of these two lines and
compare this with that obtained from line emissivity profiles
\Citep[Figures 6 and 7 of][]{wu01}.  Here we assume that the height of the
maximum emissivity region of a line coincides with the height determined
from its bulk velocity. From \Citet{wu01}, the ratios are found to be
$\sim3$ and $\sim5$ for white dwarf of mass 1.0~M$_\sun$ and 0.5~M$_\sun$
respectively. The ratio of the heights of these two lines in AM~Her is
$\sim 2$ indicating a mass closer to 1~M$_\sun$.

Using the mean emissivity over the region of maximum emissivity of the four
emission lines \ion{Fe}{26}, \ion{Fe}{25}(r), \ion{S}{16} and \ion{Mg}{12}
taken from APED and adopting a distance of 91~pc for AM~Her, we obtain an
average VEM of 2.54$\times10^{54}$~cm$^{-3}$.  Using this VEM and the
values of kT$_{sh}$ and $h$ listed in Table~\ref{tab-shock}, we determine
the electron density, n$^{sh}_e$ at the shock and the radius of the shock
region, r$_{sh}$, and list them in Table~\ref{tab-shock}.

In  polars, the accreting matter is assumed to be streaming along the
magnetic poles.  The bulk motion velocity ($v$) of this matter will be seen
from different angles along the line of sight due to the rotation of the
white dwarf. The apparent velocity of accreting matter at any orbital phase
$\Phi$ is given by $v_\Phi = v \cos\theta$, where, $\theta$ is the angle
between the magnetic axis and line of sight, and  is related to the
inclination angle $i$, magnetic obliquity $\beta$ and $\Phi$ as

\begin{equation}
\cos \theta\ =\ \cos i\ \cos \beta\ -\ \sin i\ \sin\beta\ \cos(\Phi +
\pi/2)
\label{eqn-polar}
\end{equation}

Assuming $i = 35^\circ\pm5^\circ$ and $\beta$ = $58^\circ \pm 5^\circ$
\Citep{brainerd85}, the observed modulation in the velocity of \ion{Fe}{26}
line can be well explained by Equation~\ref{eqn-polar} with the emitting
region moving at a velocity of $\sim\! 1100$ km~s$^{-1}$ accreting on to a
single pole.  This velocity is close to the shock velocity of an accreting
white dwarf of mass in the range of 0.6--0.7~$M_\sun$.
Thus, we interpret the observed variation in the amplitude of line shifts with
phase as due to the aspect effect according to Equation~\ref{eqn-polar}.
Similar phase dependent line velocity variation of few hundred km~s$^{-1}$
are reported in several FUV lines too
\Citep{gansicke98,mauche98,hutchings02}.  Also, observations with Low
Energy Transmission Grating onboard \Chandra X-ray observatory showed that
the emission line components are somewhat broader  than the instrumental
width, and the authors attribute this to the continuously changing angle
between the accretion column and the observer \Citep{burwitz01}.

\placetable{tab-lines}

The apparent  lack of any velocity shift in the fluorescent line is in
accordance with the standard emission models where this line is believed to
be due to reflection of the hard X-rays from the surface of the white
dwarf. The equivalent width of Fe fluorescent line ($\sim 150$ eV) combined
with an N$_H$ smaller than 10$^{+21}$ cm$^{-2}$ also supports this idea
\Citep{ezuka99}.

In the above discussion we assume that the observed velocity shifts are
real and not an artefact. This can be judged from Figure~\ref{fig-felines}
that clearly shows the varying line center of Fe XXVI and Fe XXV.  Also the
shifts in the line center values obtained from the overlapping phase bins
(Figure~\ref{fig-vel_shift}) clearly show a sinusoidal trend modulated with
the orbital period. The one artefact that might get introduced into our
analysis is wrong identification of a particular line as the wavelength
region around 1.6-2.0 $\AA$ is filled with many DES and other lines. Though
the resolution of Chandra HEG data does not allow us to fully resolve
individual He-like lines of Fe, and wrong identification of a line is
plausible, the effect of this is only to change the final velocity value
but not the modulation.

\section{ Conclusions }
\label{sec-concl}

We have presented here an analysis of high resolution \chandra observations
of a prototype polar AM~Her. The analysis of the average X-ray spectrum
shows that it is well fitted by a multi-temperature, partially absorbed
plasma emission APEC models. A mass of 1.15$\pm$0.05 M$_\sun$ is derived
based on the best fit of multi-temperature plasma model to the average
X-ray spectrum of AM~Her.  The multi-temperature nature of the plasma is
further confirmed by analysis of individual emission lines.  The plasma
diagnostics based on line-ratios of helium-triplets and Fe-L shell ions
suggests a dense ($> 2\times10^{12}$ cm$^{-3}$) plasma. The helium-like
triplets of O and Ne give a temperature of $>$2~MK, whereas the ratio of
hydrogen-like line to helium-like line of Fe suggests a temperature of
12~keV.  We have constructed the DEM distribution of a AM~Her, which is the
first time that such an exercise has been carried out for a polar. The
constructed DEM distribution using individual line fluxes shows that the
DEM is continuous with a possible minimum near $\log T = 6.8$.

From the phase resolved spectroscopy, we have found a possible evidence for
bulk motion of the ionized material in the accretion column of AM~Her. The
velocity of \ion{Fe}{26} line is modulated as a function of the orbital
phase consistent with a single pole accretion. The \ion{Fe}{26} ions from
the hottest plasma show a maximum velocity shift of $\sim$1100 km~s$^{-1}$,
that is close to the shock velocity expected for a $\sim$0.6~M$_\sun$ white
dwarf.  The velocity shifts are observed to be much smaller for
\ion{Fe}{25(r)}, and the hydrogen-like lines of \ion{S}{16} and
\ion{Mg}{12}. The variation in the velocity shifts of different ions is
used to calculate the height of the accretion column from the white dwarf
surface from where the line emission originates. An mass of
$\sim$1.0~M$_\sun$ for the mass of the white dwarf is indicated by using
the heights of \ion{Fe}{26} and \ion{Fe}{25} line emitting regions
estimated by the observed velocities of these lines. The multi-temperature
mass estimate hence, gives an upper limit for the mass of the white dwarf
owing to the bias introduced by using single temperature flow in the
accretion column. The density and temperature structure in the accretion
column are also derived.  No velocity shift is seen in the fluorescent Fe
line, consistent with its origin due to reflection from the white dwarf
surface.  Higher resolution observations are required to detect the radial
velocity modulation in this line.

\acknowledgments

We would like to thank the referee for her/his comments helping in
improving the paper.  This research has made use of the \Chandra Data
Archive (CDA), part of the \Chandra X-ray Observatory Center (CXC) is
operated for NASA by the Smithsonian Astrophysical Observatory. VRR is
pleased to acknowledge partial support from Kanwal Rekhi fellowship. The
Gaussian convolution program used in the analysis is kindly supplied by
David, P. Huenemoerder.

\bibliographystyle{apj1b}

%%%%%%%%%%%%%%%%% Tables  %%%%%%%%%%%%%%%%%%%%%%%%%%
%%\clearpage

\begin{deluxetable}{cccccc}
	\tablecaption{Spectral parameters for the best fit 4-T APEC model to
	the average spectrum of AM~Her. The VEM is calculated for a distance of
	91 pc.}
\tablewidth{0pc}
%\tabletypesize{\scriptsize}
\tablehead{
\colhead{APEC} &
\colhead{kT} &
\colhead{log(T)} &
\colhead{$\chi^2$} &
\colhead{Dof} &
\colhead{VEM} \\

\colhead{} &
\colhead{(keV)} &
\colhead{(K)}  && &
\colhead{($\times10^{53}$cm$^{-3})$}
}

\startdata
1 & 0.13$^{+0.03}_{-0.08}$  & 6.17$^{+0.10}_{-0.39}$& 1747 & 2767 & 1.28$^{+2.55}_{-0.57}$ \\
2 & 0.65$^{+0.08}_{-0.08}$	& 6.88$^{+0.05}_{-0.06}$& 1716 & 2765 & 0.25$^{+0.38}_{-0.09}$ \\
3 & 2.7$^{+0.6}_{-0.4}$	    & 7.50$^{+0.08}_{-0.08}$ & 1654 & 2763 & 3.36$^{+5.19}_{-1.26}$ \\
4 & 15.6$^{+5.2}_{-3.1}$	& 8.26$^{+0.12}_{-0.10}$ & 1624 & 2761 & 25.4$^{+25.8}_{-3.1}$ \\

\enddata
	\label{tab-apec}
\end{deluxetable}

%\clearpage
\begin{deluxetable}{lcccc}
	\tablecaption{Measured fluxes, equivalent widths and Ionic column
	densities of the emission
	lines identified in the HETG spectrum of AM Her.}

\tablewidth{0pc}
\tabletypesize{\small}
\tablehead{
\colhead{Ion} &
\colhead{Wavelength} &
\colhead{Energy} &
\colhead{Flux} &
\colhead{Equivalent Width}\\
%\colhead{n$_H\ ^\dagger$}\\

\colhead{} &
\colhead{$\mathrm{(\AA)}$} &
\colhead{$(keV)$} &
%\colhead{$photon\ cm^{-2}\ $s^{-1}$)$} &
\colhead{$(10^{-5}\ photon\ cm^{-2}\ s^{-1})$} &
\colhead{$(eV)$}
%\colhead{(cm$^{-2}$)}
}
\startdata
Fe XXVI             &$ ~1.781^{+0.001}_{-0.002}$  &
$6.956^{-0.001}_{+0.009}$ & $9.3^{+1.8 }_{-1.3 }$ & 173 \\ % & 2.6E18\\
Fe XXV (r)              &$ ~1.853^{+0.001}_{-0.001}$	 &
$6.691^{+0.002}_{-0.005}$ & $ 7.6^{+0.9 }_{-0.8}$ & 104 \\ % &  3E18\\
Fe I $K_\alpha$		&$ ~1.937^{+0.002}_{-0.001}$ & $6.402^{+0.006}_{-0.002}$ & $ 9.3^{+1.6 }_{-1.0 }$ & 151 \\
S  XVI           	&$ ~4.735^{+0.002}_{-0.001}$ &
$2.619^{+0.001}_{-0.001}$ & $ 2.6^{+0.6 }_{-0.5 }$ &  17.9 \\ % & 5.9E17\\
S  XV 				& $~5.038^{+0.001}_{-0.016}$ &
$2.461^{+0.001}_{-0.008}$ & $ 0.8^{+0.3}_{-0.4}$ &   5.2 \\ % & 2E17\\
Si XIV          	&$ ~6.184^{+0.001}_{-0.001}$ &
$2.005^{+0.001}_{-0.001}$ & $ 2.5^{+0.3 }_{-0.3 }$ &   12.2 \\ % & 4.6E17\\
Mg XII				&$ ~8.424^{+0.001}_{-0.001}$ &
$1.472^{+0.001}_{-0.001}$ & $ 1.2^{+0.2 }_{-0.2 }$ &  3.9 \\ % & 2.8E17\\
Fe XXIV				&$10.625^{+0.004}_{-0.001}$& $1.167^{+0.001}_{-0.001}$
& $ 1.2^{+0.2 }_{-0.2 }$ &  3.9\\
Fe XXIV				&$11.181^{+0.084}_{-0.002}$& $1.109^{+0.008}_{-0.001}$ & $ 1.3^{+0.3 }_{-0.4 }$  &  2.8 \\
Fe XXIV				&$11.432^{+0.033}_{-0.144}$& $1.085^{+0.003}_{-0.014}$ & $ 0.8^{+0.3 }_{-0.4}$ &  1.7 \\
Fe XXII$^2$		& 11.78 & 1.055 &  $< 0.3 $    &  \\
Fe XXII$^2$		& 11.92 & 1.040 &  $< 0.6 $    &  \\
Ne X $^1$	&$12.139^{+0.055}_{-0.002}$& $1.022^{+0.005}_{-0.001}$ & $
2.8^{+0.5}_{-0.5}$ &  5.5 \\ % & 3.1E17 \\
Ne IX  (r)     		&$13.452^{+0.136}_{-0.001}$& $0.922^{+0.009}_{-0.001}$
& $ 1.5^{+0.7 }_{-0.4 }$ &   2.5 \\ % & 2.3E17\\
Ne IX  (i)     		&$13.544^{+0.225}_{-0.031}$& $0.916^{+0.015}_{-0.002}$ & $ 0.8^{+0.7}_{-0.3}$ &  1.3 \\
Ne IX  (f)$^2$  	&$13.544^{+0.225}_{-0.031}$& $0.905^{+0.000}_{-0.000}$ & $ <0.6$ &       \\

Fe XVII        		&$15.012^{+0.022}_{-0.091}$& $0.826^{+0.001}_{-0.005}$
& $ 1.0^{+0.7}_{-0.6} $ &   1.5 \\ % & 7E17 \\
Fe XIII        		&$16.021^{+0.851}_{-0.851}$& $0.774^{+0.041}_{-0.041}$ & $ 1.2^{+0.9}_{-0.6} $ &   2.2 \\
Fe XVII        		&$17.056^{+0.065}_{-0.065}$& $0.727^{+0.004}_{-0.004}$ & $ 1.8^{+1.1}_{-0.9} $ &   5.6 \\
Fe XVII            	&17.100                  & 0.725 & $ 0.6^{+0.9}_{-0.6} $ &       \\
O  VIII        		&$18.966^{+0.193}_{-0.193}$& $0.654^{+0.001}_{-0.001}$
& $ 9.8^{+2.5}_{-2.5} $ &  19.8 \\ % & 1.4E18\\
O  VII (r)     		&$21.603^{+0.130}_{-0.548}$& $0.574^{+0.003}_{-0.005}$
& $14.7^{+ 4.6}_{- 4.1}$   &  18.2 \\ % & 2.3E18\\
O  VII (i)    		&$21.812^{+0.013}_{-0.001}$& $0.569^{+0.001}_{-0.001}$ & $ 7.5^{+4.2}_{-3.0}$  &  9.6 \\
O  VII (f)$^2$  	&$22.064$& $ 0.562$
& $<$3.6  &  \\
\enddata
	\label{tab-avgl}
\tablenotetext{1}{Blend of Ne X and Fe XXII lines}
\tablenotetext{2}{Flux is only an upper limit obtained by freezing the
line center.}
\end{deluxetable}

%\clearpage

%%%%%%%%%%%%%%  shock parameters
\begin{deluxetable}{lccccc}
	\tablecaption{Measured line velocity shifts of emission lines and
	corresponding heights of emitting regions of
	AM~Her  \label{tab-lines}}
	\tablehead{
		 \multicolumn{1}{c}{Line Id} &
		 \colhead{V$^{(1)}_{avg}$} &
		 \colhead{V$^{(2)}_{\Phi,\ const}$} &
		 \colhead{V$^{(2)}_{\Phi,\ mod}$} &
		 \colhead{$V^{(3)}_{l}$} &
		 \colhead{Height} \\
		&
		\colhead{(km~s$^{-1}$)} &
		\colhead{(km~s$^{-1}$)} &
		\colhead{(km~s$^{-1}$)} &
		\colhead{(km~s$^{-1}$)} &
		\colhead{(km)}
	}
	\startdata
	\ion{Fe}{26} &   430$\pm130$ & 220$\pm$ 26 &
	790$\pm$40 & 565$\pm$195 & 932$\pm$110  \\

	\ion{Fe}{25}(r)&  850$\pm90$   &  660$\pm$ 70 &
	 145$\pm$110  & 770$\pm$75 &529$\pm$213    \\

	\ion{S}{16}     & 510$\pm80$     & 545$\pm$140 &
	110$\pm$120 & 500$\pm$160 & 315$\pm$220 \\

	\ion{Mg}{12}   & 190$^{+140}_{-270}$ & 240$\pm$160 &
	180$\pm$140 & 280$\pm$195 &105$\pm$130  \\
	\enddata
	\tablenotetext{1}{The error bars are based on Cash statistic with 90\%
	confidence for a single parameter.}
	\tablenotetext{2}{Sinusoid+constant fit to the velocity shifts obtained
	from non-overlapping phase bins.}
	\tablenotetext{3}{A constant fit  to the velocity shifts obtained
	from non-overlapping phase bins.}
\end{deluxetable}

%\clearpage
%%%%%%%%%%%%%%%%%%%%%%%%%
%%% Table shock
%%%%%%%%%%%%%%%%%%%%%%%%%%%
	\begin{deluxetable}{ccccccc}
	\tablecaption{Parameters of shock in accretion column of AM~Her
	\label{tab-shock}}
	\tablehead{
		\colhead{M$_{wd}$} &
		\colhead{R$_{wd}$} &
		\colhead{$v_{sh}$} &
		\colhead{kT$_{sh}$}&
		\colhead{$h$}   &
		\colhead{n$^{sh}_e$} &
		\colhead{r$_{sh}$} \\

		\colhead{(M$_\sun$)} &
		\colhead{(R$_\sun$)} &
		\colhead{(km~s$^{-1}$)}	&
		\colhead{(keV)}     &
		\colhead{(km)} &
		\colhead{(10$^{15}$ cm$^{-3}$)} &
		\colhead{(km)}
	}
	\startdata
		0.39   &  0.016 &      765     &   11.3  &  394 &    3.4 & 2556 \\
		0.50   &  0.014 &      922     &   16.4  &  691 &    2.3 & 2806 \\
		0.60   &  0.013 &     1066     &   21.9  & 1067 &    1.7 & 3017 \\
  		0.75   &  0.011 &     1290     &   32.1  & 1889 &    1.2 & 3318 \\
  		0.91   &  0.009 &     1555     &   46.7  & 3312 &    0.8 & 3643 \\
		\enddata
	\end{deluxetable}

%\clearpage

%%\clearpage

%\clearpage

%\clearpage

%\clearpage

\end{document}